\documentclass{emulateapj}
\usepackage{amssymb}
\usepackage{amsmath}
\usepackage{epstopdf}
\usepackage{natbib}
\usepackage{xcolor}

\usepackage[colorlinks, citecolor=blue]{hyperref}
\usepackage[all]{hypcap}
\usepackage[flushleft]{threeparttable}

\usepackage{mwe,tikz}

\begin{document}

\title{CDF-S XT1 and XT2: white dwarf tidal disruption events by intermediate-mass black holes? }
\author{
Zong-Kai Peng$^{1,2}$, Yi-Si Yang$^{1,2}$,
Rong-Feng Shen$^{3}$,Ling-Jun Wang$^{4}$, Jing-Hang Zou$^{5}$, Bin-Bin Zhang$^{1,2}$}

\begin{abstract}

Recently two fast X-ray transients (XT1 and XT2) have been reported from the search in the 
Chandra Deep Field (CDF) data. Each transient shows an initial plateau lasting around 
hundreds to thousands seconds followed by a rapid decay in the light curve.
In particular, CDF-S XT2 is found to be associated with a galaxy at redshift $z$ = 0.738 and 
was explained as a counterpart of a binary neutron-star merger event.
In this paper, motivated by the short duration and decay slopes of the two transients, we 
consider an alternative interpretation in which both events are accretion-driven flares from 
tidal disruption of white dwarfs by intermediate-mass black holes. We derive a theoretical 
model of the accretion rate history, and find that it fits the observed X-ray light curves 
well. The extremely super-Eddington peak luminosity of XT2 can be explained by the beaming 
effect of the system, likely in the form of a jet.

\end{abstract}


\affil{\altaffilmark{1}School of Astronomy and Space Science, Nanjing
University, Nanjing 210093, China; bbzhang@nju.edu.cn} 
\affil{\altaffilmark{2}Key Laboratory of Modern Astronomy and Astrophysics (Nanjing University), Ministry of Education, China} 
\affil{\altaffilmark{3}School of Physics \& Astronomy, Sun Yat-Sen University, Zhuhai 519082, China; shenrf3@mail.sysu.edu.cn}
\affil{\altaffilmark{4} National Astronomical Observatories, Chinese Academy of Sciences, Beijing 100012, China }
\affil{\altaffilmark{5}Department of Space Sciences and Astronomy, Hebei Normal University, Shijiazhuang 050024, China}

\section{Introduction}

\label{sec:Intro}

The physical process of the tidal disruption event (TDE) which happens when a star is 
disrupted by a black hole (BH), has been investigated by many authors \citep{Hil1975,Lac1982,Car1983,Ree1988,Eva1989}. The radiation of TDEs ranges from optical to X-ray energy bands. \citet{Dai2018} proposed a unified model of TDE and pointed out that the 
different emission may be caused by the different viewing angles. \citet{Dai2018} also 
studied the disk dynamics of the TDE through general relativistic radiation 
magnetohydrodynamic simulations, and proposed that a jet can be produced. If a jet does exist in a TDE, the observations would be subject to the beaming effect. Some recent TDEs 
observations indeed suggest the existence of a jet component \citep{Blo2011,Cen2012,Bro2015}. 
When a main-sequence star is disrupted by a BH, the process will last for years. However the 
time scale will be much shorter if the disrupted object is a compact star such as a white 
dwarf (WD) \citep{krolik11,haas12,Lod2012,Kaw2018}. In the case of WD-involved TDE, the BH 
mass can not exceed $\sim \rm few\times 10^{5}M_{\odot}$ (hence a stellar-mass BH or an 
intermediate-mass BH), otherwise the WD is swallowed as a whole and there will be no observed emission \citep{Cla2012,Kaw2018}. 

Since the Laser Interferometer Gravitational-Wave Observatory (LIGO) and Virgo discovered the first Gravitational-Wave (GW) event GW150914 \citep{Abb2016a,Abb2016b}, stellar-mass ($\sim 10~M_{\odot}$) BHs are realized to be common in the universe. Meanwhile, mounting evidence 
shows that supermassive BHs ($>10^{6}M_{\odot}$) exist at the centers of most galaxies \citep{Kor1995}. The origin of intermediate-mass BHs (IMBHs), on the other hand, remains an 
open question and some recent observations suggest they do exist in the centers of dwarf 
galaxies or star clusters \citep{Far2009,Lin2018,chiling18}.

Recently two fast X-ray transients from the 7-Ms Chandra Deep Field-Source survey, namely CDF-S XT1 \citep{Bau2017} and CDF-S XT2 \citep{Xue2019}, were reported. The CDF-S XT1 seems to be associated with a faint galaxy without any spectroscopic redshift measurement\footnote{\cite{Bau2017} pointed out a photometric redshift of 2.23 with large uncertainties.}. \citet{Xue2019} pointed out that CDF-S XT2 was
associated with a galaxy at redshift 0.738 and lies in its outskirts with a moderate offset of $0.44\pm 0.25$ arcsec.

The light curve of CDF-S XT1 shows a rise within $\sim 100~\rm s$ to the peak flux of $F_{0.3-10~\rm KeV} \approx 5\times 10^{-12}~\rm erg~\rm s^{-1}~\rm cm^{-2}$, then a power law decay in hours with a slope of $-1.53 \pm 0.27$ \citep{Bau2017}. The light curve of CDF-S XT2 shows a long plateau, and then a sudden break at $\sim 3000~\rm s$. Its peak luminosity between $0.3$ and $10~\rm KeV$ is $\approx 3\times 10^{45}~\rm erg~\rm s^{-1}$. 
For XT2, \citet{Xue2019} reported the best-fitting power-law indices of $-0.14^{+0.03}_{-0.03}$ and $-2.16^{+0.26}_{-0.29}$ before and after the break. 
 
Without establishing a confirmed redshift (hence luminosity), there were several theoretical models to explain CDF-S XT1: an orphan afterglow of a short gamma-ray burst; a low-luminosity gamma-ray burst with a large redshift; or a TDE of IMBH with WD \citep{Bau2017}. For CDF-S XT2, by considering its luminosity, host galaxy offset and event rate, \citet{Xue2019} pointed out that it most likely originated from a magnetar which was formed after a binary neutron-star (NS) merger event, a possibility considered by \citet{Xiao2019, Sun2019, LHJ2019} as well. While other possibilities are not entirely ruled out. 

The late temporal decay slope of CDF-S XT2 is steeper than the canonical value -5/3 of a stellar TDE's debris mass fallback rate \citep{Ree1988}. In addition, we noticed that the time scale of CDF-S XT2 is much shorter than those of stellar TDEs but it fits well the scenario of a WD-involved TDE. In this Letter we explore the tidal encountering process of a WD and an IMBH and its subsequent accretion as a possible origin of these two CDF-S transients. A similar scenario is considered by \citet{She2019}, which proposed a model that a WD is tidally stripped by an IMBH to explain two fast, ultraluminous X-ray bursts found by \citet{Irw2016}.

We describe the model in Section \ref{sec:model}, paying particular attention to the role of the disk viscous accretion in shaping the light curve. The model prediction is then fitted to the observed light curve and the results are presented in Section \ref{sec:fit}. The discussion and conclusions are summarized in Section \ref{sec:con}.

\section{The Model}

\label{sec:model}

We aim to interpret CDF-S XT2 and XT1 as accretion transients resulted from a TDE of a WD encountering an IMBH. We consider that the WD approaches the IMBH on a parabolic orbit.
When the {WD} reaches the tidal radius \citep{Hil1975,Ree1988,Can1990,Koc1994}
\begin{equation}
R_{\rm t}\simeq R_{*} \left(\frac{M_{\rm BH}}{M_{*}}\right)^{1/3},
\label{eq1}
\end{equation}
the surface material will be disrupted by BH, where $M_{\rm BH}$ is the mass of BH, and $R_{*}$ and $M_{*}$ are the radius and mass of the WD, respectively. We consider both cases of a full disruption and a partial disruption (`stripping'). For the tidal stripping case, a factor of $2$ is introduced in front of $R_*$ in Eq. (\ref{eq1}) \citep{She2019}.

An analytical solution of WD mass-radius relation was derived by \cite{Nau1972}. By fitting to a simple power law, we find it can be roughly approximated 
{by $R_*/R_{\odot} \simeq 0.0078 m_*^{-2/3}$} for the mass range $0.2 < m_* < 1.2$, where $m_{*} = M_{*}/M_{\odot}$ with $M_{\odot}$ being solar mass and $R_{\odot}$ being solar radius. This approximation serves to ease our analysis of parameter dependence later. Substituting $R_*$ in Eq. (\ref{eq1}), we have 
\begin{equation}
R_t \simeq 18 \left(\frac{M_{\odot}}{M_*} \right) \left(\frac{10^3 M_{\odot}}{M_{\rm BH}}\right)^{2/3} R_S,
\end{equation}
where $R_{S}$ is the Schwarzschild radius of the BH. 

Similar to the stellar TDE case, the bound portion of the disrupted WD material falls back to the disruption site with a mass rate history of
\begin{equation}	\label{eq:mdotfb}
\dot{m}_{\rm fb}(t) = \frac{\Delta m}{t_{\rm fb}}\frac{(n-1)}{n}f(t/t_{\rm fb}),
\end{equation}
where
\begin{equation}	\label{eq:fx}
f(x)=
\begin{cases}
1,~~~~~~~~for&x\leq 1,\\
x^{-n},~~~~for&x>1.
\end{cases}
\end{equation}
Here $\Delta m$ is the total bound mass {and $n = 5/3$ for typical TDEs} . The time scale over which the bulk of the debris stream {falls back is}
\begin{equation}	\label{eq:tfb}
t_{\rm fb}= \frac{\pi R_t^3}{\sqrt{2G M_{\rm BH} R_*^3}} \simeq 77 \left(\frac{M_{\rm BH}}{10^3 M_{\odot}}\right)^{1/2} \left(\frac{M_{\odot}}{M_*}\right)^2~ \mbox{s},
\end{equation}
where the last step uses Eq. (\ref{eq1}) and the WD's mass-radius relation. 
After a time of $t_{\rm fb}$, the fallback rate drops as $\propto t^{-5/3}$. In the above calculation, $t = 0$ is set to the epoch when the first parcel of disrupted material falls back to the disruption site.

The returned mass cannot be digested promptly by the BH. {After forming a disk, it will swirl} inward within
the disk. This process can be accounted for by a viscous accretion time scale $t_{\rm acc}(R_{\rm t})$; it is the time
that each parcel of mass has to spend before reaching the BH. It depends on the tidal radius
$R_{\rm t}$ -- where the disrupted material firstly returns to, and on the physical regimes of the disk as in
\begin{equation}	\label{eq:tacc}
\begin{split}
t_{\rm acc}(R_{\rm t}) &= \frac{1}{\alpha}\sqrt{\frac{R_{\rm t}^{3}}{GM}} \left(\frac{H}{R}\right)^{-2} \\
 & \simeq 23~m_{*}^{-1} \left(\frac{0.1}{\alpha}\right) \left(\frac{H}{R}\right)^{-2}~{\rm s},
\end{split}
\end{equation}
where $\alpha$ is the viscosity parameter \citep{Sha1973,Fra2002}, and $H/R$ is the disk thickness-to-radius ratio. The last step of Eq. (\ref{eq:tacc}) utilized Eq. (\ref{eq1}) and the WD mass-radius relation. It shows that $t_{\rm acc}$ is insensitive to the BH mass $M$, though $H/R$ may contain a subtle $M$-dependence. 

If the disk is in the radiatively efficient and geometrically thin regime (i.e., Shakura \& Sunyaev disk), $H/R$ $\approx 0.01~\dot{m}^{1/5}(\alpha M_{\rm 4})^{-1/10}r^{1/20}$, where $\dot{m}$ and $r$ are the accretion rate and disk radius \textit{normalized} by $L_{\rm Edd}/(0.1c^{2})$ and $R_{\rm S}$, respectively \citep{kato08}.
If the disk is in the advective-cooling dominated and geometrically thick regime (i.e., slim disk; \cite{Abr1988}), $H/R$ is approximately unity. The borderline between the two regimes is $\dot{m}\sim r/10$. 
In any case, $t_{\rm acc}(R_t)$ ranges from $\sim 10^{2}$ s to $\sim 10^{5}$ s given $0.01\le H/R\le 1$.

With a mass supplied from fallback and a drain due to accretion, the global temporal evolution of the disk can be written as:
\begin{equation}	\label{eq:dmdt}
\frac{dm_{\rm d}}{dt} = \dot{m}_{\rm fb}(t)-\dot{m}_{\rm acc}(t),
\end{equation}
and the accretion rate can be approximately expressed as:
\begin{equation}
\dot{m}_{\rm acc}(t) = \frac{m_{\rm d}}{t_{\rm acc}(R_{\rm t})}.
\end{equation}

A general solution of $m_{\rm d}(t)$ to equation (\ref{eq:dmdt}) can be obtained in a time-integrated form \citep{Kum2008}. Since $t_{\rm acc}$ can be roughly regarded as a constant as all disrupted material returns to the same radius, the accretion rate history can be solved as:
\begin{equation}
\dot{m}_{\rm acc}(t) = \frac{1}{t_{\rm acc}}\int_{0}^{t}\dot{m}_{\rm fb}(t') \exp\left[-\frac{(t-t')}{t_{\rm acc}}\right] dt'.
\label{eq:9}
\end{equation}
{Combining equations (\ref{eq:mdotfb}-\ref{eq:fx}) and (\ref{eq:9}) , the solution can be rewritten as:}
\begin{equation} \label{eq:mdot}
\dot{m}_{\rm acc}(t) = \frac{\Delta m}{t_{\rm acc}} A(t, t_{\rm fb}, t_{\rm acc}, n),
\end{equation}
where, with given parameters $t_{\rm fb}$, $t_{\rm acc}$ and $n$, the function
\begin{equation}
\begin{split}
A(t, t_{\rm fb}, t_{\rm acc}, n) = \frac{n-1}{n} ~~~~~~~~~~~~~~~~~~~~~~~~~~~~~~~~\\ 
\times \int_{0}^{\frac{t}{t_{\rm fb}}} f(t'/t_{\rm fb}) \exp\left[-\frac{(t-t^{'})}{t_{\rm acc}}\right] d(t'/t_{\rm fb})
\end{split}
\end{equation} 
contains all the temporal shape information of $\dot{m}_{\rm acc}(t)$ and its peak value is $\sim~1$.

\begin{figure}[h]
\begin{center}
\includegraphics[width=0.4\textwidth,angle=0]{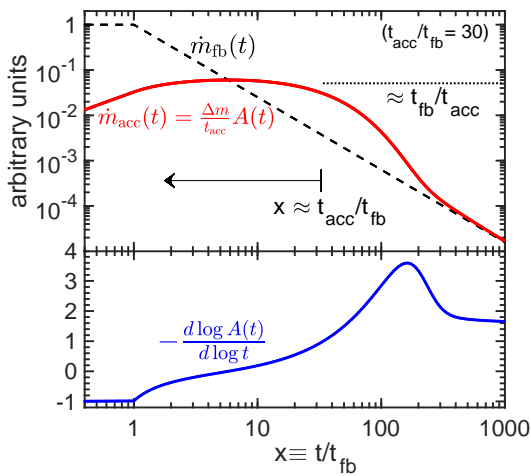} %
\end{center}
\caption{{An example of the accretion rate history (red solid line) calculated from equation (\ref{eq:mdot}) for a given mass fallback rate history (black dashed line; equation \ref{eq:mdotfb}). The lower panel plots the instantaneous temporal slope of the accretion rate.}} 
\label{fig:Ay}
\end{figure}

Figure \ref{fig:Ay} shows an example of the shape of the accretion rate history $\dot{m}_{\rm acc}(t)$. Compared with the short-duration, fast decaying mass supply curve $\dot{m}_{\rm fb}(t)$, the accretion rate shows a `slowed' plateau, followed by a steep drop toward the decaying tail of the supply rate. The duration of the plateau is $\approx t_{\rm acc}$, and the level of the plateau, or the peak accretion rate, is $\approx \Delta m / t_{\rm acc}$. The post-plateau drop is not possessed of an asymptotic slope (not until $t \gg t_{\rm acc}$), unlike the case of the spin-down power rate of a young pulsar. However, its instantaneous slope (shown in the lower panel) is certainly steeper than $n$, and is $\approx$ 2 - 3. As will be shown below, such values are consistent with the observed slope CDF-S XT1 and XT2.

Introducing a constant radiative efficiency $\eta$, we calculate the bolometric luminosity light curve from the accretion rate history:
\begin{equation}
L(t)=\eta \dot{m}_{\rm acc}(t)c^{2}=\eta c^{2}\frac{\Delta m}{t_{\rm acc}} A(t, t_{\rm fb}, t_{\rm acc}, n).
\end{equation}
For CDF-S XT2, its luminosity is $\sim 10^{45}$~erg~s$^{-1}$ and the duration of the plateau of the light curve is $\approx 3000$ s. So we have
\begin{equation}
\begin{split}
L(t) = 6\times 10^{44} \left(\frac{\eta}{0.1}\right) \left(\frac{\Delta m}{10^{-5} M_{\odot}}\right) ~~~~~~~~~~~~\\
\times \left(\frac{3000~\rm s}{t_{\rm acc}}\right) A(t, t_{\rm fb}, t_{\rm acc}, n) ~{\rm erg~s^{-1}}.
\end{split}
\end{equation}

\section{The Fit}
\label{sec:fit}

We fit the light curve of CDF-S XT2 using Eq. (13) with three free parameters: $t_{\rm acc}$, $t_{\rm fb}$ and $\delta =\log(\eta \Delta m /M_{\odot})$, the last of which is introduced in recognition of the degeneracy between $\Delta m$ and $\eta$; $n$ is fixed to 5/3. Using a Monte Carlo (MC) fitting tool developed by ourselves \citep{Zha2015}, we get the best fitting parameters listed in Table \ref{tab:para}. 

\begin{figure*} \centering
\begin{tikzpicture}[ every node/.style={anchor=south west,inner sep=0pt}, x=1mm, y=1mm,] 
 \node (fig1) at (0,0)
 {\includegraphics[width=7.5in]{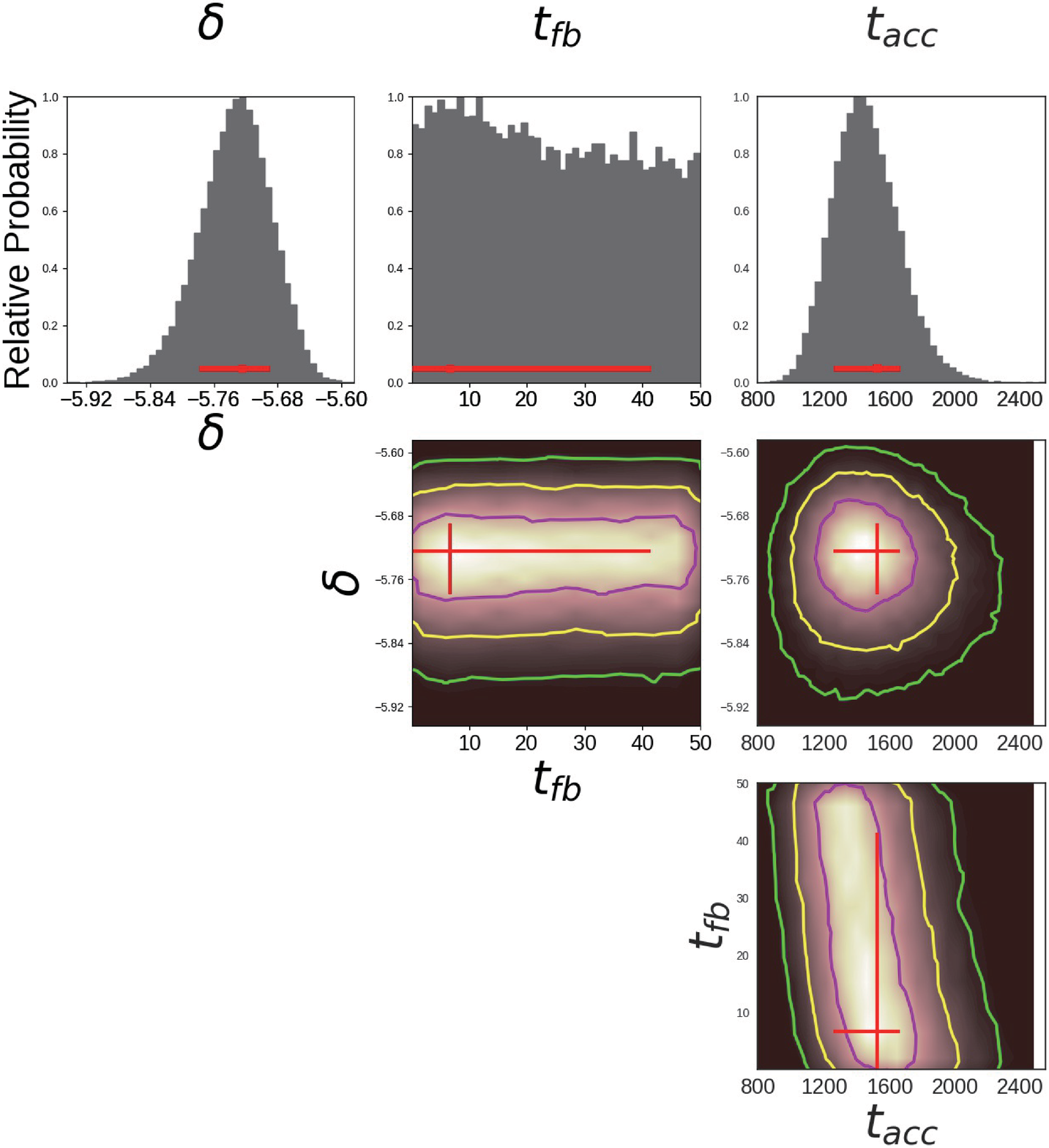}};
\node (fig2) at (10,5)
 {\includegraphics[width=3.5in]{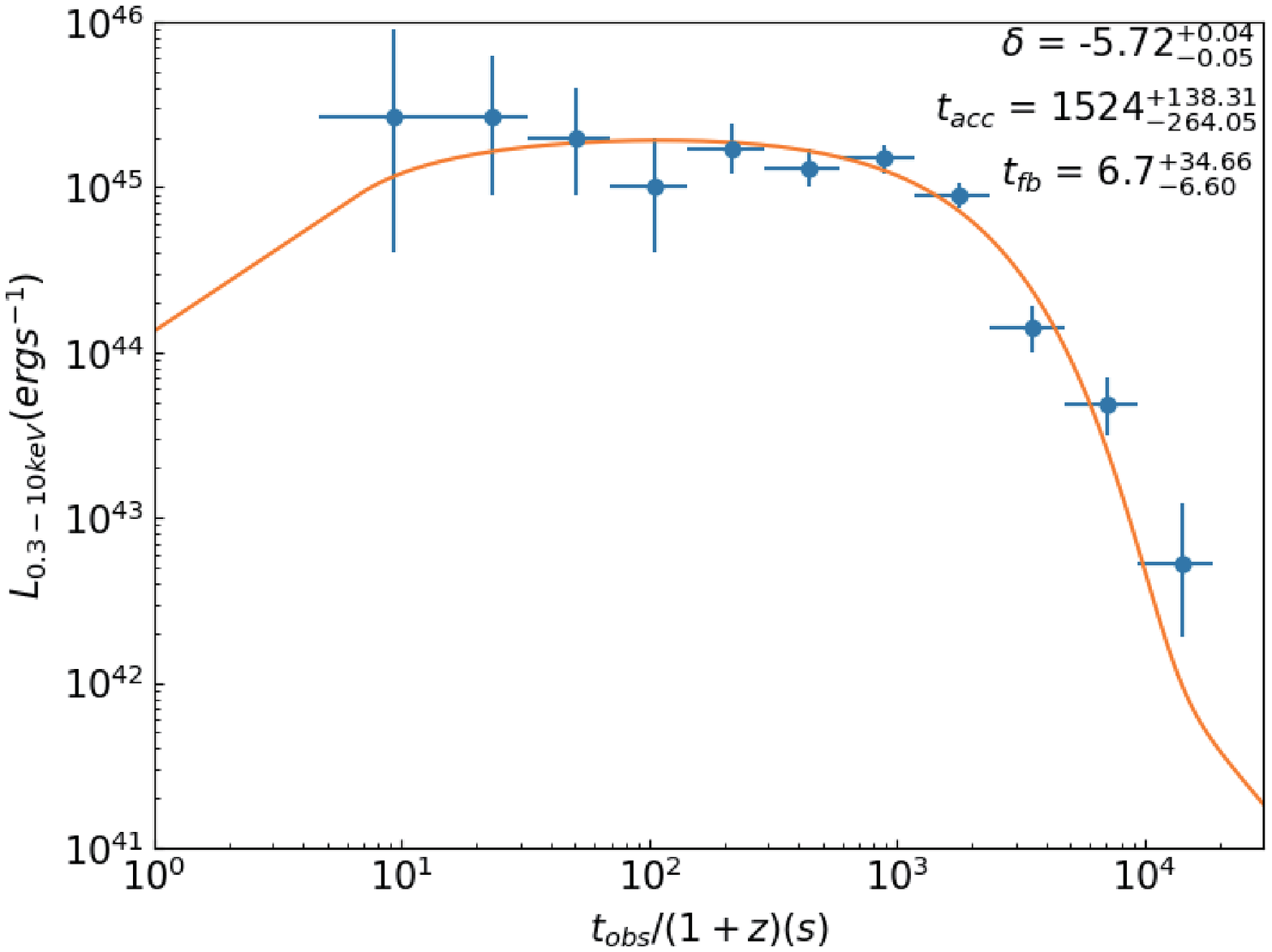}};
\end{tikzpicture}
 \caption{{}{Fitting our TDE model to the observed data of CDF-S XT2. The left-bottom plot shows the best-fit modeled light curves (solid line) over-plot on observed on the data points (filled circles). Right-top corner plot shows the constraints of the three best-fit parameters. }}
\label{fig:XT2-P}
\end{figure*}

The goodness of the fit meets the condition $\chi ^{2}$ = 11.71 $<\chi ^{2}_{\alpha=0.05,DOF=8}=15.51$, suggesting that the fit is acceptable at 0.05 significance level. The parameter constraints as well as the best-fit model curves are presented in Figure \ref{fig:XT2-P}. While $\delta$ and $t_{\rm acc}$ are constrained reasonably well, we found $t_{\rm fb}$ is not sensitive in our fit but is consistent with a small value in the parameter space. A small $t_{\rm fb}$ might be caused by a large WD mass (cf. equation \ref{eq:tfb}). 

The latest two data points of XT2 seem to show a slightly shallower decay than the model predicted. We find that if we adjust the value of $n$ to 1.5 or 1.3, these two data points fit better and the overall goodness of fitting is improved. Although some numerical simulations of partial disruptions \citep{Gui2013,coughlin19} or those taking into account some realistically evolved stellar structures \citep{Law2019, Gol2019} tend to find $n$'s that are steeper than 5/3, so far in the TDE literature there has been no finding that $n < 5/3$. On the other hand, the simulations mentioned above are all about disruptions of normal stars. For disruptions of compact stars like WDs considered here, could it be $n < 5/3$? It is an interesting question for future numerical exploration.

\begin{figure*} \centering
\begin{tikzpicture}[ every node/.style={anchor=south west,inner sep=0pt}, x=1mm, y=1mm,] 
 \node (fig1) at (0,0)
 {\includegraphics[width=7.5in]{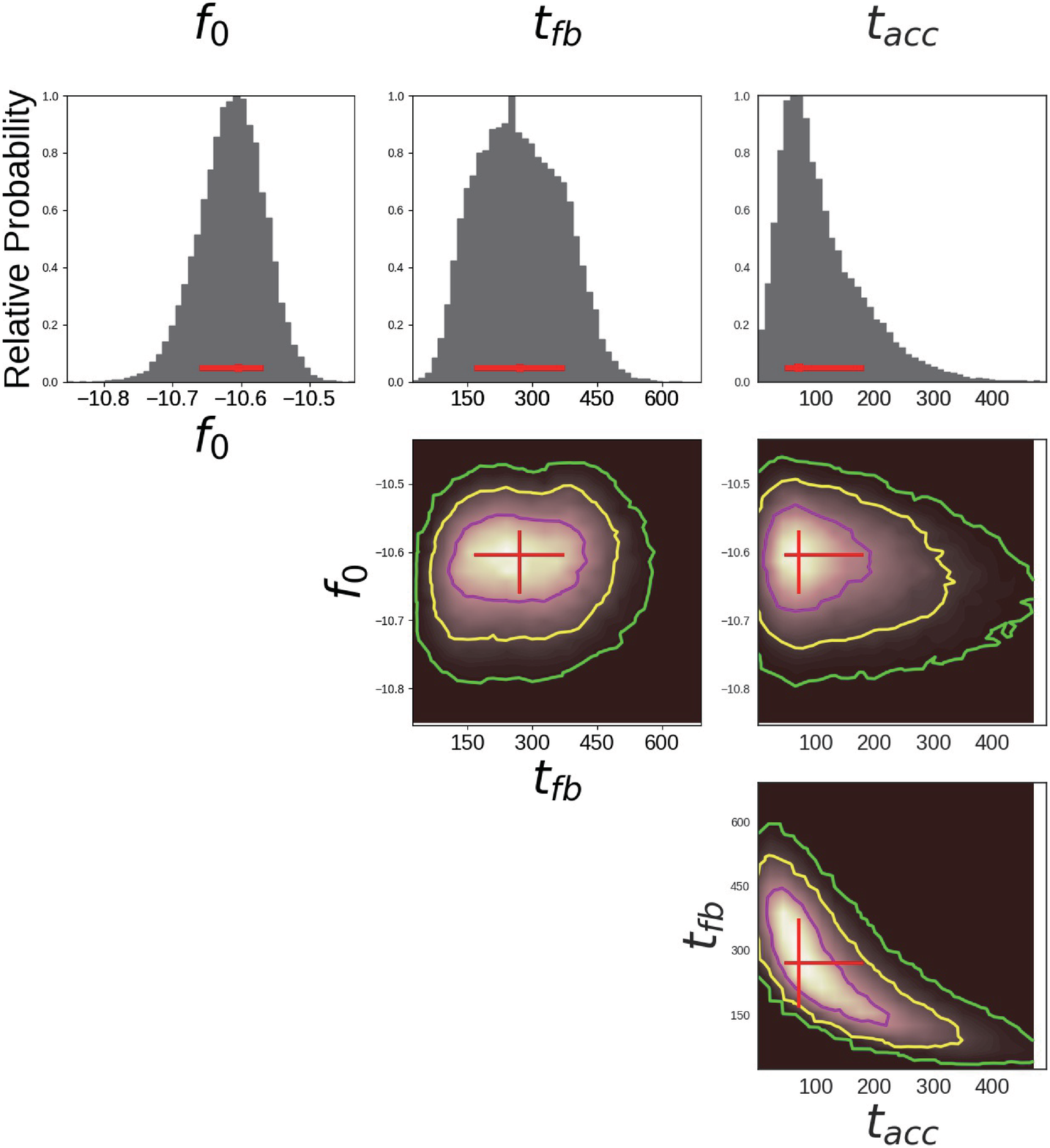}};
\node (fig2) at (10,5)
 {\includegraphics[width=3.5in]{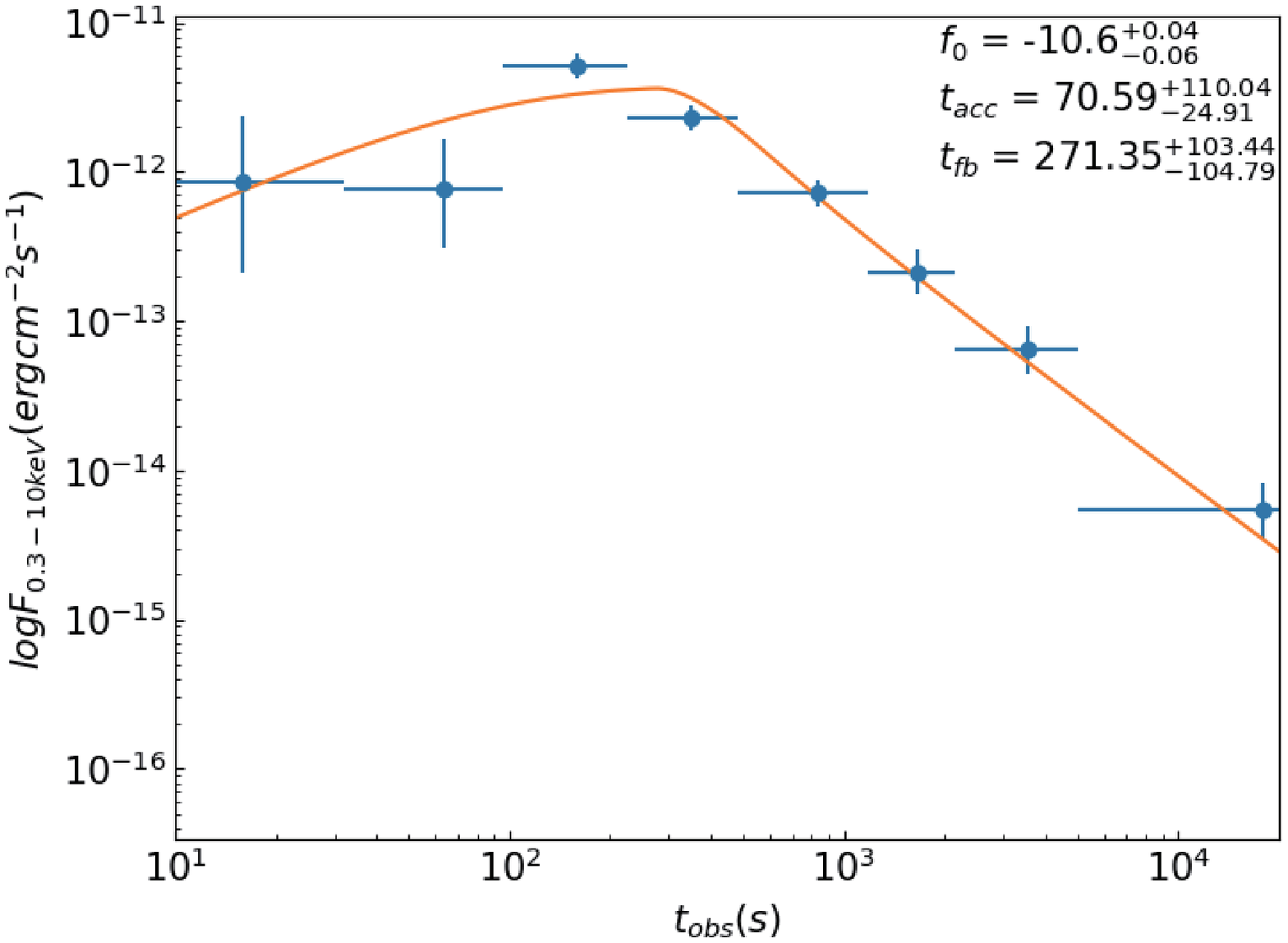}};
\end{tikzpicture}
 \caption{{}{Same as Fig. 3 but for the fit of CDF-S XT1. The fitting model is described in Eq. (14). The $f_{0}$ is logarithmic.} }
\label{fig:XT1-P}
\end{figure*}

Similarly, we fit the light curve of CDF-S XT1 with the same approach. We use one free parameter $f_{0}$ to account for $\eta$, $\Delta m$ and the unknown redshift $z$. So the observed flux light curve can be written as: 
\begin{equation}
F(t) = f_{0} \times \left(\frac{100~\mbox{s}}{t_{\rm acc}}\right) A(t, t_{\rm fb}, t_{\rm acc}, n) ~{\rm erg~cm^{-2}~s^{-1}}.
\end{equation}
Figure \ref{fig:XT1-P} shows the fitting to the light curve of XT1 and the parameter corner.

The XT1 fit yields a $\chi^2=11.48$ which is $< \chi^2_{\alpha=0.05,DOF=6}=12.59$, indicating the fit is still acceptable at 0.05 significance level. The second and third data points contribute the most of residuals after the fit, which we interpret as being due to some early fast variabilities of the jet luminosity.

Compared with XT2, a shorter duration of the plateau of XT1 leads to a shorter accretion timescale. The best fit of XT1 gives $t_{\rm fb} \simeq 271$ s and $t_{\rm acc} \simeq 71$ s. Those values are consistent with the estimation using equations (\ref{eq:tfb}) and (\ref{eq:tacc}). The relatively small $t_{\rm acc}$ indicates $H/R \sim 1$ ({thus,} a slim disk), which is consistent with a super-Eddington accretion case \citep{Abr1988, Dot2011}.

\section{Discussion and Conclusions}
\label{sec:con}

Motivated by the short time scales and high luminosities of two recently discovered X-ray transients CDF-S XT1 and XT2, we calculate the light curve of an accretion-driven transient when a WD is tidally disrupted by an intermediate-mass BH. {We find that the model fits well to both events. A similar model has been used to explain some ultraluminous X-ray bursts \citep{She2019}. 
By introducing a viscous accretion time scale, this model has more flexibility in explaining the temporal behavior of light curve, than the conventional stellar TDE model. For example, the late-time slope of XT2 is $\approx -2.1$ which is steeper than those of other typical TDEs \citep{Lac1982, Ree1988, Li2002}, but it can be well accounted for using Eqs. (13 - 14). The mass of the central black hole in our model is also flexible but recent observations (e.g, 3XMM J215022.4-055108; Lin et al. 2018) suggest IMBHs may be common in off-centre star clusters which may fit in well with the central object in our model.

For a WD approaching the BH, three possible types of the orbit are permitted, namely elliptical, parabolic and hyperbolic \citep{Li2002, Kob2004}. If the orbit is elliptical, a periodicity of the light curve is expected, which is not observed. This may suggest that the orbit is parabolic or hyperbolic for XT1 and XT2.

Comparing the best-fit results of CDF-S XT1 with CDF-S XT2 in Table \ref{tab:para}, we notice that $t_{\rm fb}$ of CDF-S XT2 is shorter, but its $t_{\rm acc}$ is longer. This suggests that, if the BH masses in the two systems are the same, the WD in the case of XT2 might be heavier (i.e., more compact), so its disruption radius is closer to the BH (cf. eq. \ref{eq:tfb}). The longer $t_{\rm acc}$ in XT2 is likely due to a lower disk thickness ratio $H/R$ because the latter carries the most sensitive parameter dependence in eq. (\ref{eq:tacc}). A lower $H/R$ in turn might be caused by a lower Eddington-normalized accretion rate. This could suggest that XT2 was a tidal stripping event, because the total bound mass $\Delta m$ can be much smaller than that of a full disruption event.

The isotropic-equivalent peak luminosity of CDF-S XT2, $\approx3\times 10^{45}~$erg~s$^{-1}$, is extremely high, which raises the possibility of a relativistically beamed emission. Three relativistic jetted TDE candidates have been discovered so far: Sw J1644+57 \citep{Blo2011,Bur2011}, Sw J2058+0516 \citep{Cen2012,pasham15} and Sw J1112-8238 \citep{Bro2015}.}
The {2-D} model of \citet{Dai2018} {suggests} that the X-ray emission of TDEs may be caused by jets through the Blandfold-Znajek process. Similarly, the {WD-IMBH tidal disruption / stripping discussed here may produce a jet as well.} 

\citet{Xue2019} estimated the CDF-S XT2-like event rate density to be $1.3^{+2.8}_{-1.1}\times 10^{4}~$Gpc$^{-3}$yr$^{-1}$. For CDF-S XT1, \citet{Bau2017} estimated a large range of the event rate, $\sim$ $10^3$ Gpc$^{-3}$ yr$^{-1}$ for $z=0.5$ to $\sim$ 1 Gpc$^{-3}$ yr$^{-1}$ for $z=3$.

Theoretically, the rates of WD disruptions are very uncertain. For IMBHs in globular clusters (GCs), \cite{baumgardt04} estimated a total stellar TDE rate of $\sim 10^{-7}$ yr$^{-1}$ via N-body simulations of GCs with an initial central BH mass of $10^3 M_{\odot}$. Among the disrupted stars, $\sim 15\%$ are WDs, giving a rate of $\sim 1.5 \times 10^{-8}$ yr$^{-1}$. Adopting a number density 34 Mpc$^{-3}$ of GCs, this gives a volumetric rate of $R_{\rm IMBH-WD} \sim 500$ yr$^{-1}$ Gpc$^{-3}$ \citep{haas12,shcher13}. Recently, \cite{Fra2018} semi-analytically calculated the evolution of a population of GCs in a galaxy and found a rate of WD TDE $\sim 10^{-5}$ yr$^{-1}$ per galaxy; combining this with GC population's dependence on redshift and galaxy types, their results show a present-day volumetric rate of $R_{\rm IMBH-WD} \sim$ 10 yr$^{-1}$ Gpc$^{-3}$. 

For IMBHs in dwarf galaxies, \cite{macleod14} calculated a rate of $\sim 10^{-6}$ yr$^{-1}$ per IMBH for WD disruptions via the loss-cone dynamics for the BH mass range of $10^4 - 10^5 M_{\odot}$. This rate is $\sim 30$ times lower than that of main-sequence stellar TDEs by SMBHs (e.g., \cite{stone16}). Assuming a number density of dwarf galaxies $\sim 10^{7}$ Gpc$^{-3}$ \citep{shcher13,macleod14} and an occupation fraction $f_{\rm IMBH}$ of IMBHs in dwarf galaxies, then the volumetric rate is $R_{\rm IMBH-WD} \sim 10 f_{\rm IMBH}$ yr$^{-1}$ Gpc$^{-3}$.

Compared with WD-IMBH TDEs, a similar case which happens more common is the tidal disruption / stripping of main-sequrence (MSs) stars by IMBHs. Indeed, as was shown in \citet{Fra2018}, the event rate of MS-IMBH TDEs is about 30 times higher than that of WD-IMBH TDEs in most galaxies. However, the time scale ($\sim$years) of MS-IMBH TDEs is much longer than that of WD-IMBH TDEs ($\sim$hours) \citep{Che2018}, which might disguise themselves as persisting sources, thus hindering the identification of their transient nature. For disruptions of evolved stars like giants, the corresponding time scales are even ($\sim$10 times) longer \citep{Mac2012}. From the time scale consideration, XT1 and XT2 are unlikely to be MS TDEs. \citet{Che2018} predicted a detection rate of 20 MS-IMBH TDEs per year by Zwicky Transient Factory (ZTF), and 0.03 yr$^{-1}$ by Chandra. \citet{Lin2018} reported a MS-IMBH TDE candidate 3XMM J215022.4-055108, from which they inferred the event rate of 3XMM J215022.4-055108 like TDEs to be $\sim$10 Gpc$^{-3}$ yr$^{-1}$. This low detection rate may suggest that most of the MS-IMBH TDEs might have been missed due to their slow-evolution disguise.

Future detection of more similar events might either rule out the model presented here for those events or clear up our current ignorance about the IMBH demographics.

BBZ thank the hospitality of X. Liu during the visit at Xinjiang Observatory and the support by the National Key R\&D Program of China under grant number 2018YFA0404602.  This work is also supported by the National Key Research and Development Program of China (2018YFA0404204) and NSFC-11833003. R.-F.S. is supported by NSFC grant 11673078. L.J.W. acknowledges the support from the National Program on Key Research and Development Project of China (grant 2016YFA0400801). We thank Bing Zhang and Ye Li for helpful discussions and the anonymous referee for helpful suggestions.

\clearpage

\begin{table*}
\begin{threeparttable}
\begin{center}
\caption{Best fitting parameters from MCMC code.}
\label{tab:para}
\begin{tabular}{cccccccccccc}
\hline
\hline
Parameters & $\delta ^{*}$ & $f_{0} ^{**}$ & $t_{acc}$(s) & $t_{fb}$(s) & chisq/dof \\

\hline

 XT1 &$\cdots $ &$-10.6^{+0.04}_{-0.06}$ & $70.59^{+110.04}_{-24.91}$ & $271.35^{+103.44}_{-104.79}$ & 11.48/6.0 \\
 \\

 XT2 & $-5.72^{+0.04}_{-0.05}$&$\cdots$ & $1523.90^{+138.31}_{-264.05}$ & $6.70^{+34.66}_{-6.60}$ & 11.71/8.0 \\

\hline
\hline
\end{tabular}
\begin{tablenotes}
\item[] $^{*}$Because the stripped mass $\Delta m$ and $\eta$ are degenerated with each other, we combined them into one single parameter $\delta =\log(\eta * \frac{\Delta m}{M_{\odot}})$.
\item[] $^{**}$The redshift of CDF-S XT1 is not given, so we combined redshift $z$, $\eta$ and $\Delta m$ into one parameter $f_{0}$.
\end{tablenotes}
\end{center}
\end{threeparttable}
\end{table*}


\end{document}